\documentclass[preprint,preprintnumbers]{revtex4}

\usepackage{graphicx}

\usepackage{amssymb}

\usepackage{mathrsfs}

\begin{document}

\title{Topological Insulators  and  Superconductors - A Curved Space Approach }

\author{D. Schmeltzer}

\affiliation{Physics Department, City College of the City University of New York,  
New York, New York 10031, USA}

\pacs{}

\begin{abstract}

\noindent 
The  method of the space dependent basis is applied to study  electronic spinors in a crystal.  The crystal in the momentum space is described by the Brillouine zone which might contains obstructions or degeneracies for which requires different gauges for different regions.  The electronic bands are classified according to their topology. The connection and curvature determines   the physical properties which are clasified according to the  topological invariants. 
We apply this method to the Topological Insulators, Topological Superconductors, Persistent Currents  
in coupled rings and photoemission for a curved  crystal-face boundary.

\vspace{0.15 in}

\noindent
Keywords:Topological Insulators, connection,curvature, curved space, Topological Invariants

\end{abstract}

\maketitle


\vspace{0.2 in}
\noindent
\textbf{1.-Introduction}

\vspace{0.2 in}

\noindent
 One of the important ideas  in Condensed Matter Physics is the concept  of topological order \cite{Volkov,Haldane,Golterman,Kreutz,Thouless,Berry,Mele,Kane, Bellissard,davidSpinorbit, More,ZhangField,Hasan,Chao,Zhangnew,Simon, Prodan,rings,genus}. Insulators   with a single Dirac cone  which lies in a gap  such  as    $Bi_{2}Se_{3}$, $Bi_{2}Te_{3}$  and $Bi_{1-x}Sb_{x}$ \cite{Volkov,Hasan}  represent the  experimental realization of  Topological Insulators ($T.I$). 
 At the surface   of the three dimensional  Topological Insulator ($T.I.$), one  obtains a two dimensional  metallic surface  characterized  by  an odd number of chiral excitations,  due to  Kramers theorem,   electrons are  protected against backscattering  \cite{balatsky} and localization  \cite{Hai,davidT}. When time reversal is broken  localization effects are observed \cite{Ando}. The surface  physics  has been realized  in     $CdTe/HgTe/CdTe$  quantum wells.   The  quantized  spin-Hall effect has been proposed    \cite{Haldane} and  observed by \cite{Zhangnew,Wu} and recently the Anomalous Hall effect has been measured \cite{Takahashi}.
The  spin resolved photoemission  \cite{Hasan,Nature,FanZ} has been used to identify the surface states.  Topological superconductors  and their identification through the Majorana Fermions have  been observed \cite{Alicea}.
In order to study physical properties in the Brilouine Zone ($B.Z.$)    we need  to  use   the concept of parallel  transport since  the spinors might rotate   in   $B.Z.$
The Brillouin zone contains obstructions, degeneracies and  therefore the gauge transformation between  different   different regions is needed.  This behavior is studied  with the help of  the  \textbf{ connection} (the  vector potential in the momentum  space) \cite{Nakahara}    which is similar to  the     parallel  transport  on    curved surfaces.  The derivative (external)\cite{Nakahara}  of the  \textbf{connection}  defines the \textbf{curvature}    strength   which   measures  the obstructions in the Brillouin Zone.  According to the symmetry  involved  (time reversal symmetry, parity inversion, mirror  symmetry or  charge conjugation)
the  eigenvectors   satisfy  certain constraint equations.  The solutions  of  the constraint symmetry gives rise to   specific gauge symmetry for the connection in the momentum space.  This gauge symmetry is  used to  compute the electromagnetic response with the coefficients which characterizes topological invariants .    The electromagnetic  (magnetoelectric) response is characterized by  the \textbf{second Chern number}..  The interplay of \textbf{Topological Insulators} ($T.I.$) and \textbf{Superconductivity}  gives  rise to \textbf{Majorana Fermions} \cite{Alicea}.

\noindent
The plan of this paper is as folows: In Sec.2. the method of paralel transport in the momentum space is introduced. Sec.3. is devoted to the studies of topological invariants which can be obtained using external fields to measur the responce of the system. In Sec.4.1 we consider the topological invariants for superconductors. In Sec.4.2 we derive the topological  response  which is obtained from sound waves.
In Sec.5 we show derive the equation of motion in the momentum space.
Sec.6. is devoted to the computation of topological invariants in two space dimensions using the mapping to four space dimensions. In Sec.7. we study the topology of two coupled rings in the presence of Majorana fermion  and compute the persistent current.
In Sec.8 we study the effects of the topology of curved surfaces on the Photoemission.
In Sec.9 we present our conclusion.

\vspace{0.2 in}

\noindent
 \textbf{2. -The method of parallel  transport in momentum space}

\vspace{0.2 in}

\noindent 
In this section I will develop the method of Topological Insulators based on the ideas of parallel   transport in a curved space. The space is represented by the $B.Z.$
The  spinors in the  presence of  the  spin-orbit interaction \cite{davidSpinorbit}  wary from point to point in $B.Z.$. This demands the use of   parallel transport  $\langle \psi(\vec{k})| \bigtriangledown_{a}\psi(\vec{k})\rangle =0$ which is defined in terms of the spinors spinors $|U_{\alpha}(\vec{k})\rangle$ ($\alpha$ is the band-spin index). The parallel transport is defined in  terms  of the connection  $ A^{\alpha,\beta}_{a}(\vec{k})\equiv  \langle U_{\alpha}(\vec{k})|i\partial_{a}|  U_{\beta}(\vec{k})\rangle=\int\,d^{d}x U^{*}_{\alpha,k}(x)( x^{a}) U_{\beta,k}(x)$ and the  \textbf{curvature} $[X_{a},X_{b}]=F_{a,b}$  for the coordinates. This allows to introduce   the first and second \textbf{chern character}   $ch_{1}= \frac{i}{2\pi}\mathbf{Tr}\Big( F_{a,b}(\vec{k})\Big)$,  $ch_{2}= \frac{-1}{32 \pi^2} \epsilon^{a,b,c,d}\mathbf{Tr}\Big( F_{a,b}(\vec{k}) F_{c,d}(\vec{k})\Big)$  which are   given in terms of the \textbf{covariant}  coordinate  (or polarization)  $X_{a}=x_{a}+A_{a}(\vec{k})$ ( $ x_{a}$ is the coordinate in the momentum space   $ x_{a}=i\partial_{k^{a}}$) . 
This tools are essential for studies of  Topological Insulators and Superconductors  which are gaped phases of Fermionic systems which exhibit topological protected boundary modes to arbitrary deformation as long discrete symmetry such as \textbf{ time reversal, particle hole and chiral symmetry } are respected. Due to the  symmetries the  Hamiltonian  $h(\vec{k})$ is invariant at the time invariant point $\vec{k}=\vec{\Gamma}$  which obey $Th( \vec{\Gamma})T^{-1}=h(-\vec{\Gamma})$ (time reversal with $T^2=-1$) or charge conjugation $Ch( \vec{\Gamma})C^{-1}=-h^{*}(-\vec{\Gamma})$  ($C^2=-1$). The product of the charge conjugation (particle-hole) with the time reversal allows to define the \textbf{unitary chiral symmetry} which holds in the entire  $B.Z.$. As a result we have $10$   symmetry classes. 
For the case that the inversion symmetry  $P$ and the  time reversal  $T$ symmetry hold, it has been proposed \cite{Kane,Fu} that computing $\langle U^{(-)}(\vec{\Gamma})|P|U^{(-)}(\vec{\Gamma})\rangle= \mathbf{Sgn}M(\Gamma)$  determine the topological invariant index $Z_{2}$, $I=\prod_{i}\mathbf{Sgn}M(\Gamma_{i})$ .
The real challenge is to relate the index $I$ to  the \textbf{quantized  electromagnetic response} \cite{ZhangField,Zhangnew} .

\vspace{0.1 in}

\noindent  
 \textbf{2.1. The space dependent basis in the B.Z.}

\vspace{0.1 in}
\noindent  
  The seminal work  of \cite{Bellissard} on the Integer Quantum Hall opened the door  to study disorder as a problem in a curved space, this work has been further developed in the Mathematical literature by  \cite{Prodan} and his collaborators.  In an early paper we have realized that due to   the  spin-orbit interaction \cite{davidSpinorbit}  the  spinors  wary from point to point in $B.Z.$.  
\noindent
 Using the formal  language of  space dependent basis  we  introduce   the \textbf{ tangent} vector.  In the local frame we have  a set of vectors $ e_{a}(\vec{x})$ which are  related to the cartesian coordinates, 
  $ e_{a}(\vec{x})=e^{\mu}_{a}(\vec{x}) \partial_{\mu}$,  $\vec{V}(\vec{x})=\sum_{a}V^{a}(\vec{x})e_{a}(\vec{x})$
 
\noindent
  For translational invariant systems  we have in the B.Z.  the Bloch spinors which are  a  momentum  dependent basis  $|\vec{k}\otimes \eta_{\alpha}(\vec{k})\rangle \equiv  |U_{\alpha}(\vec{k})\rangle$  (when orbitals are included  $|\vec{k}\otimes \sigma_{i}(\vec{k})\otimes  \tau_{s}( \vec{k})\rangle \equiv |U_{\alpha}(\vec{k})\rangle$ ).

\noindent
 In the presence of a \textbf{non translational} invariant  potential $V(x)$  we   replace
 $ |U_{\alpha}(\vec{k})\rangle$ by,
  $|\tilde{U}_{\alpha}(\vec{k})\rangle=\int\,dq\sum_{\beta} Z_{\alpha,\vec{k}  }^{\beta,\vec{k}+\vec{q}} |U_{\beta}(\vec{k}+\vec{q})\rangle$.
Any Bloch Spinor can $ |\psi(\vec{k})\rangle$  be represented in terms of the momentum  dependent  basis   $ |U_{\alpha}(\vec{k})\rangle$  (or $|\tilde{U}_{\alpha}(\vec{k})\rangle$),
$|\psi(\vec{k})\rangle=\sum_{\alpha} C^{\alpha}(\vec{k})  |U_{\alpha}(\vec{k})\rangle$

\noindent
 In Quantum Mechanics  we  have the following matrix element  for the momentum derivative $\partial_{a}=\partial_{k^{a}}$.
   In real space- $\langle x| \partial_{a} |y \rangle= -i  X^a \delta(x-y)$  , $X^a$ is the coordinate operator.  In momentum space we have , -$\langle k| \partial_{a} |p \rangle=  \partial_{a} \delta(k-p)$

 \noindent
We borrow  the following concepts from Differential Geometry \cite{Nakahara}
$d=\partial_{a}dk^{a}$  is the exterior  derivative which allow to introduce $id|U_{\alpha}(\vec{k})\rangle$.  $\hat{A}_{a}(\vec{k})$ is the spin connection and  $\hat{F}(\vec{k})$ is the curvature operator.

\noindent
 \textbf{The  spin connection  operator}:
 $\hat{A}_{a}(\vec{k})\equiv \sum_{\alpha,\beta}A^{\alpha,\beta}_{a}(\vec{k})|U_{\alpha}(\vec{k})\rangle \langle U_{\beta}(\vec{k})|$
with the  matrix element:
$-i A^{\alpha,\beta}_{a}(\vec{k})\equiv \langle U_{\alpha}(\vec{k})|\partial_{a}|  U_{\beta}(\vec{k})\rangle=\int\,dx U^{*}_{\alpha,k}(x)(-i X^{a}) U_{\beta,k}(x)$

\noindent 
\textbf{The curvature operator} :
$\hat{F}_{a,b}(\vec{k})  \equiv \sum_{\alpha,\beta} F^{(\alpha,\beta)}_{a,b}(\vec{k}) |U_{\alpha}(\vec{k})\rangle \langle U_{\beta}(\vec{k})| $,
 with the matrix elements:
$ F^{(\alpha,\beta)}_{a,b}(\vec{k})=\partial_{a} A^{\alpha,\beta}_{b}(\vec{k})-\partial_{b} A^{\alpha,\beta}_{a}(\vec{k}) +i[\partial_{a} A^{\alpha,\gamma}_{b}(\vec{k}),\partial_{b} A^{\gamma,\beta}_{a}(\vec{k})]$


\noindent 
\textbf{The derivative of   the operator $\hat{f}= \sum_{\alpha,\beta}f^{\alpha,\beta}(\vec{k}) |U_{\alpha}(\vec{k})\rangle \langle U_{\beta}(\vec{k})|$} 
\begin{eqnarray}
&&\partial_{a}\hat{f}=\partial_{a}(f^{\alpha,\beta}(\vec{k})) |U_{\alpha}(\vec{k})\rangle \langle U_{\beta}(\vec{k})|+f^{\alpha,\beta}(\vec{k})\partial_{a}[  |U_{\alpha}(\vec{k})\rangle \langle U_{\alpha}(\vec{k})]\nonumber\\&&
=\partial_{a}(f^{\alpha,\beta}(\vec{k})) |U_{\alpha}(\vec{k})\rangle \langle U_{\beta}(\vec{k})|  +(i)\Big[\hat{X}^{a},\hat{f}\Big]\nonumber\\&&
\hat{X}^{a}=[X^{a}]^{\alpha,\beta}(\vec{k}) |U_{\alpha}(\vec{k})\rangle \langle U_{\beta}(\vec{k})|\nonumber\\&&
\end{eqnarray}

\noindent
 \textbf{The covariant derivative for the spinors} 

\noindent
$|\nabla_{a}\psi(\vec{k})\rangle=\sum_{\alpha}\Big[ \partial_{a}  C_{\alpha}(\vec{k})    -i A^{\alpha,\gamma}_{a}(\vec{k}) C_{\gamma}(\vec{k})  \Big] |U_{\alpha}(\vec{k})\rangle$

\noindent
\textbf{Parallel transport $\langle\psi(\vec{k}) |\nabla_{a}\psi(\vec{k})\rangle=0$}

\noindent
 $| \Psi(\vec{k},P)\rangle=|\psi(\vec{k})\rangle P \Big[e^{-i\int_{-\infty}^{\vec{k}}\,dk'^{a} A_{a}(\vec{k}')}\Big]$, \hspace{0.1 in}
$|\psi(\vec{p})\rangle= P \Big[e^{-i\int_{\vec{p}}^{\vec{k}}\,dk'^{a} A_{a}(\vec{k}')}|\psi(\vec{k})\rangle$

\noindent
\noindent 
The physics   of  electrons in  a periodic crystal  is  determined   by the  eigenvectors (spinors)    $|U_{n}(\vec{k})\rangle$ ($n$ is the band-spin index) behavior   in  the  \textbf{ Brillouin Zone}  $\vec{k}\in T^{d}$ (torus in a $d$ dimensional  momentum space). This behavior  is similar   to  the parallel  transport  of a vector  around  a  curve.  We need to  find  the way the eigenvectors  change under transport in the Brillouin Zone  \cite{davidSpinorbit,Blount,Zak}.
The topological properties are encoded into  the   connection  $ \mathcal{A}_{i}$ (the vector potential in the momentum space) which measures  the changes of   $|U_{n}(\vec{k})\rangle$ when it  is transported  in the Brillouin Zone. The changes are   given by:

\noindent 
 $i d|U_{n}(\vec{k})\rangle-\Gamma^{m}_{i;n} dk^{i}| U_{m}(\vec{k})\rangle=0 $  (an index  which appears twice  implies  a summation ).  The matrix  $ \Gamma^{m}_{i;n}$ is given by
$ \Gamma^{m}_{i;n}\equiv
i\langle U_{n}(\vec{k})|\partial_{k^{i}}| U_{m}(\vec{k})\rangle=  \mathcal{A}^{(n,m)}_{i}(\vec{k})$  where   $\mathcal {A}$ is   the\textbf{ connection}.
Applying twice the (\textbf{exterior}) derivative  we define the \textbf{ curvature}   $ \mathcal{F}=d(d|U_{n}(\vec{k})\rangle $    ( see eqs.$ 7.145  -  7.145$  , Nakahara (2008) page 285 \cite{Nakahara}) and find :

\noindent
 $\mathcal{F}=d(d|U_{n}(\vec{k})\rangle)=\Big[[\partial_{j}\Gamma^{l}_{i;n}(\vec{k})+\Gamma^{m}_{j;n}(\vec{k})\Gamma^{l}_{i;m}(\vec{k})]dk^{i}\wedge dk^{j}\Big]|U_{l}(\vec{k})\rangle=\frac{1}{2} [F_{i,j}]_{n,l}dk^{i}\wedge dk^{j}|U_{l}(\vec{k})\rangle$ ;  ( the symbol $\wedge$ represents the wedge product )

\noindent 
$ [F_{i,j}]_{n,m}=\partial_{i}\mathcal{A}^{(n,m)}_{j}(\vec{k})-\partial_{j}\mathcal{A}^{(n,m)}_{i}(\vec{k}) +i[\mathcal{A}^{(n,l)}_{i}(\vec{k}),\mathcal{A}^{(l,m)}_{j}(\vec{k})]$,

\noindent
 where $F_{i,j}$ is the matrix curvature with the matrix elements  $[F_{i,j}]_{n,m}$  given in terms of the commutator of the  \textbf{covariant derivative}  $\hat{R}_{i}=\hat{r}_{i}+\mathcal{A}^{(n,m)}_{i}$ , $\hat{r}_{i}=i\partial_{k^{i}}$ ; $ [\hat{R}_{i},\hat{R}_{j}]=F_{i,j}$.

\vspace{0.1 in}

\noindent
 \textbf{2.2  Observing the topology   using external sources  or disorder potential }

\vspace{0.1 in}

\noindent
 The Hamiltonian can  be express in terms of the eigenvalues.
To obtain   information about the topology, we have to transport the spinor around the B.Z.
Alternatively we  can include space dependent scalar and vector potentials 
\textbf{$a_{0}(\vec{x})$, $\vec{a}(\vec{x})$} and probe  the response.

\noindent 
 The $TI$ Hamiltonian  $ H_{0}$ with spin  half and two orbitals
 $|\vec{k}\otimes \sigma=\uparrow,\downarrow\otimes \tau=1,2\rangle$ in the presence  of  the external vector  potential $\vec{a}(\vec{x})$   and scalar potential $a_{0}(\vec{x})$ is given by:
\noindent
 $ H=H_{0}+\int \,d^d x[\rho(\vec{x})a_{0}(\vec{x})+\vec{J}(\vec{x})\cdot \vec{a}(\vec{x})]\equiv H_{0}+H^{ext.} $
\noindent The four component spinors for the Hamiltonian $H_{0}$:

\noindent 
 $ \psi(\vec{x})=\int\,d^dk e^{i\vec{k}\cdot\vec{x}} \psi(\vec{k})$;
$ \psi(\vec{k})=\sum_{s=1,2}[ C_{s}(\vec{k})U_{s}(\vec{k})+b^{\dagger}_{s}(-\vec{k})V_{s}(-\vec{k})]$.
 $U_{s}(\vec{k})$  and $ V_{s}(\vec{k})$   are four component  spinors for particles  and antiparticles.
\begin{eqnarray}
&&H^{ext.} = \int\,dx^{d} \psi^{\dagger}(\vec{x})V(\vec{x})\psi(\vec{x})\nonumber\\&&=\int\, d^{d} k \int\, d^{d} q \sum_{s,s'}\Big[C^{\dagger}_{s}(\vec{k})V(\vec{q})\langle U_{s}(\vec{k})|U_{s'}(\vec{k}+\vec{q})\rangle C_{s'}(\vec{k}+\vec{q})\Big]\nonumber\\&&
\approx \int\, d^{d} q  V(\vec{q}))(-iq^{a}) \sum_{s,s'}  \int\, d^{d} k C^{\dagger}_{s}(\vec{k}) \Big[\delta_{s,s'}i\partial_{a}+A_{a}^{(s,s')}(\vec{k})\Big]C_{s'}(\vec{k})\nonumber\\&&
i\delta_{s,s'}\partial_{a}+A_{a}^{(s,s')}(\vec{k})\equiv \hat{X}^{a}\nonumber\\&&
\Big[\hat{X}^{a},\hat{X}^{b}\Big]\equiv \hat{F}_{a,b}(\vec{k})
\end{eqnarray}
\textbf{ The surface $T.I.$}
Due to T.R.S. invariance with $ T^2=-1$ and finite chemical potential $\mu$ the integrated \textbf{ Fermi Surface curvature is $\pi$}.

\noindent
 The situation is similar to spin-orbit scattering giving rise to $ anti-localization$
This can be demonstrated using \textbf{ a diagrammatic or a Non linear sigma approach}
Due to the \textbf{spin connections } we find that the Cooperon  changes sign!
As a result the conductivity increases.
The surface Hamiltonian is given by :
$h(\vec{k} ,\vec{x}) =-\sigma_{2} k_{1}+\sigma_{1} k_{2}+V(\vec{x})$.
For a finite chemical potential $\mu>0$ we have the eigen spinors $U_{\sigma}(\vec{k})$
\begin{eqnarray}
&&\Psi_{\sigma}(\vec{k})= C(\vec{k})  U_{\sigma}(\vec{k})\nonumber\\&&
U_{\sigma=\uparrow}(\vec{k})=\frac{1}{\sqrt{2}}e^{\frac{i}{2}\chi(\vec{k})}; U_{\sigma=\downarrow}(\vec{k})=\frac{1}{\sqrt{2}} i e^{\frac{i}{2}\chi(\vec{k})}; 
\chi(-\vec{k})=\chi(\vec{k})+\pi\nonumber\\&&
\end{eqnarray}
The effect of the random potential:
\begin{eqnarray}
&&H^{ext.}=\int\,d^2x \Psi^{\dagger}(\vec{x})V(\vec{x})\Psi(\vec{x})=
\int\, d^{2} k \int\, d ^{2}q \Big[V(\vec{q})C^{\dagger}_{s}(\vec{k})\Big(U^{*}(\vec{k}),U(\vec{k}+\vec{q})\Big)C(\vec{k}+\vec{q})\Big]\nonumber\\&&
\approx \int\, d^2 q  V(\vec{q}))(-iq^{a})  \int\, d ^{2}k C^{\dagger}(\vec{k}) \Big[i\partial_{a}+A_{a}(\vec{k})\Big]C(\vec{k})\nonumber\\&&\partial_{a}
A_{a}(\vec{k})=\partial_{a}\frac{1}{2}\chi(\vec{k})\nonumber\\&&
\end{eqnarray}
As a result the multiple scattering matrix $ S$ obeys:
\begin{equation}
S(\vec{k}\rightarrow -\vec{k})=e^{i\pi}TS(\vec{k}\rightarrow -\vec{k})=-S(-\vec{k}\rightarrow \vec{k})
\label{sattering}
\end{equation}
This result is the reason  for  anti-localization which we have obtained in ref. \cite{davidT}.

\vspace{0.2 in}

\noindent
\textbf{3.Topological invariants from response theory}

\vspace{0.2 in}

\noindent
 In order to demonstrate the emergent of the topological invariant  we will consider a typical  $T.I.$   Hamiltonian  for the materials  $ Bi_{2}Se_{3}$, $ Bi_{2}Te_{3}$, $Sb_{2}Te_{3}$ . We  introduce    the  tensor product  $|\alpha\rangle\equiv|\sigma=\uparrow,\downarrow\rangle\otimes |\tau=1,2\rangle$    ($\sigma$ stands for the $ spin$  and    $\tau$  stands for the  orbitals).   A four band model   is obtained \cite{Chao}  which  can be  written in  the chiral form.
\noindent
$h(k)=I \epsilon(\vec{k})+\gamma_{1}\hat{k}_{2}+\gamma_{2}\hat{k}_{1}+ \gamma_{3}\eta\hat{k}_{3}+ \gamma_{0}M(\vec{k})$.

\noindent 
 The first term affects only the eigenvalues  and not the eigenvectors , $\eta<<1$.
 The $ \gamma$  matrices  are given as a tensor product  :$\gamma_{i}=\sigma^{i}\otimes(-1) \tau_{2}$, $i=1,2,3$ , $\gamma_{0}=I\otimes\tau_{1}$ ,$\Gamma_{5}=I\times \tau_{3}$.
 The mass  (gap)  $M(\vec{k})$ obeys $M(-\vec{k})=M(\vec{k})$ and has  points in the Brillouin where it vanishes. (On a lattice with the lattice constant $a$ we define     the Cartesian component of the  momentum  $\hat{k}_{i}=\frac{ \sin[k_{i} a]}{a}$.) 
 The Hamiltonian $h(k)$ is  diagonalized using the four eigenvectors $ |U_{s}^{(e)}(\vec{k})\rangle$, $s=1,-1$ are  the  \textbf{spin helicity} operator and $e=+,-$  represents  the particles-antiparticles energies,  $E(\vec{k})= \epsilon(\vec{k}) \pm\sqrt{k^2+M^{2}(\vec{k})}$ with the  mass   $M(\vec{k})$ which vanishes at $\vec{k}^{*}$ and   $M(\vec{\Gamma})\neq0$ :
\begin{equation}
 h(k)=\sum_{s=1,-1}\Big[E(\vec{k})|U_{s}^{(e=+)}(\vec{k})\rangle \langle U_{s}^{(e=+)}(\vec{k})| + E(\vec{k})|U_{s}^{(e=+)}(\vec{k})\rangle\langle U_{s}^{(e=+)}(\vec{k})|\Big]
\label{bas}
\end{equation}
\noindent    The Green's function operator in the  $\alpha$  basis $|\alpha\rangle\equiv |\sigma=\uparrow,\downarrow\rangle\otimes|\tau=1,2\rangle$  is given by:
\noindent
 $\hat{G}(\omega,\vec{k})= \hat{G}(\omega,\vec{k})_{\alpha,\alpha'}|\alpha\rangle \langle \alpha'|$. 
\noindent
In the \textbf{eigen vector   basis}  the Green's  function takes the form:
\begin{equation}
\tilde{G}(\omega,\vec{k})=\sum_{s=1,-1}[\frac{ |U_{s}^{(e=+)}(\vec{k})\rangle\langle U_{s}^{(e=+)}(\vec{k})|}{ \omega - E(\vec{k})+i\epsilon}+ \frac{ |U_{s}^{(e=-)}(\vec{k})\rangle \langle U_{s}^{(e=-)}(\vec{k})|}{ \omega + E(-\vec{k})+i\epsilon}]
\label{equation}
\end{equation}
\noindent The transformation from  the  $|\alpha\rangle$   basis to the eigenvector  basis  $|U_{s}^{e}(\vec{k})\rangle$  replaces the coordinate $\hat{r}_{i}=i\partial_{k}^{i}$  with  \textbf{ the covariant coordinate $\hat{R}_{i}$} \cite{davidSpinorbit}.
\noindent  In the second quantized form  the spinor operator $\Psi(\vec{r})$ is given by: 
\noindent
 $\Psi(\vec{r})=\int\,\frac{d^d k}{(2\pi)^{d}}  e^{i\vec{k}\cdot \vec{r}}\Psi(\vec{k})$,
\noindent
$\Psi(\vec{k})=\sum_{s=1,-1}\Big[C_{s}(\vec{k})U^{(+)} _{s}(\vec{k}) +  b^{\dagger}_{s}(-\vec{k})U^{(-)} _{s}(-\vec{k})\Big]$, 
\noindent
$ \bar{\Psi}(\vec{k})=\Psi^{\dagger}(\vec{k})\gamma_{0}$.
\noindent
 (It is important to stress that this representation is valid for momentum $|\vec{k}|< |\vec{k}^{*}|$, for  the region $|\vec{k}|> |\vec{k}^{*}|$  we need to choose a different representation.)The coupling of the $T.I.$  to the electromagnetic field  $\vec{a}(\vec{r},t)$ and  $ a_{0}(\vec{r},t)$ is given by the action   $S^{ext}$:
\noindent
\begin{eqnarray} 
&&S^{ext}=\int\,\frac{d^d k}{(2\pi)^{d}}\int\,\frac{d^d Q}{(2\pi)^{d}} \int\,\frac{d \omega}{2\pi} \int\,\frac{d \Omega}{2\pi}[  \bar{\Psi}(\vec{k},\omega) \gamma_{\nu}a_{\nu}(\vec{Q},\Omega)\Psi(\vec{k}+\vec{Q},\omega+\Omega)] \nonumber\\&&  \approx\int\,d^3 x\int\,dt[\sum_{\nu=0}^{3}\sum_{\mu=0}^{3}\gamma_{\nu}(\partial_{\mu}a_{\nu}(\vec{r},t))|_{\vec{r}=0,t=0)} ]\int\,\frac{d^{d}k}{(2\pi)^d}\int\,\frac{d\omega}{2\pi}
\bar{\Psi}(\vec{k},\omega)\gamma^{\mu}\hat{R}_{\mu}\Psi(\vec{k},\omega)]\nonumber\\&&
\bar{\Psi}(\vec{k},\omega)=\Psi^{\dagger}(\vec{k},\omega)\gamma_{0};  \hat{R}_{\mu=0}=i\partial_{\omega}\nonumber\\&&
\end{eqnarray}
 \noindent
We compute the  \textbf{partition function}  integrating over the \textbf{Grassman fields}  for four  space dimensions. We  find that  the effective action for  the electromagnetic fields  obtained  by \cite{Golterman} is given by, 
\noindent
 $Z=\int\prod_{s}\prod_{k}\prod_{\omega}dC_{s}(\vec{k},\omega)d C^{\dagger}_{s}(\vec{k},\omega)db_{s}(-\vec{k},\omega)d b^{\dagger}_{s}(-\vec{k},\omega) e^{i(  S^{0}+S^{ext})}=e^{i \Gamma[a_{0}(t,\vec{r}),\vec{a}(t,\vec{r})]}$

\vspace{0.1 in}

\noindent
 Using the totally antisymmetric tensor $\epsilon_{\alpha_{ 1},\alpha _{2},\alpha_{ 3},\alpha _{4},\alpha_{ 5}}$ for  four space dimensions  we find    the electromagnetic  response ,\textbf{ polarization} energy  (given in terms of the electric $\vec{E}$ and magnetic  field $\vec{B}$) : 
\noindent
$ \Gamma[a_{0}(t,\vec{r}),\vec{a}(t,\vec{r})]=\hat{C}_{2} \epsilon_{\alpha_ {1},\alpha_{ 2},\alpha_{3},\alpha_{4},\alpha_{5}}\int\,d^{4} x \int\,dt a_{\alpha_{1}}\partial_{x_{\alpha_{2}}}
a_{\alpha_{3}}\partial_{x_{\alpha_{4}}}
a_{\alpha_{5}}=\frac{\theta=\pi}{2\pi}(\frac{e^2}{hc})\int\,d^{3} x \int\,dt[\vec{E}\cdot\vec{B}]$

\vspace{0.1 in}

\noindent 
 with the quantized coefficient which dose not   break the time reversal symmetry \cite{Golterman}.
\noindent The  response   is given for     $4+1$ space time  dimensions.   \cite{Golterman,Weinberg,Nakahara,davidtop,ZhangField,Zhangnew},
\begin{equation}  
c_{2} =const.\epsilon^{0,i,j,k,l}Tr[\tilde{G}(\omega,\vec{k})\hat{R}_{0}\tilde{G}^{-1}(\vec{k})\tilde{G}(\omega,\vec{k})\hat{R}_{i}\tilde{G}^{-1}(\omega,\vec{k})\tilde{G}(\omega,\vec{k})\hat{R}_{k}\tilde{G}^{-1}(\omega,\vec{k})\tilde{G}(\omega,\vec{k})\hat{R}_{l}\tilde{G}^{-1}(\omega,\vec{k})]
\label{iint}
\end{equation}
 In the presence of  additional interactions the \textbf{Green's function  might   have   zeroe's} \cite{Gourarie,Zhong},  for such a case the system seizes to be topological. If the renormalized Green's  function has no zeroe's the topology is preserved. The renormalized Green's function  is given  in terms  of the wave function renormalization  $Z$, 
\noindent
 $\tilde{G}_{R}(\omega,\vec{k})=\sum_{s}\frac{Z^{-1} (\frac{\kappa}{\Lambda},u_{i},M)}{(\omega+i\epsilon-E_{s}(\vec{k}))}|U_{s}(\vec{k})\rangle \langle U_{s}(\vec{k})|$
\noindent
 where 
\noindent
$Z (\frac{\kappa}{\Lambda},u_{i},M)=\Big[1-\frac{\partial_{\omega}\Sigma(\frac{\omega}{\Lambda},u_{i},M)}{\partial_{\omega}}|_{\omega=\kappa}\Big]^{-1}$ 
\noindent
is the wave function renormalization.  When the wave function renormalization $Z$ is finite at  $\omega=0$  we take the limit  $\omega\rightarrow 0$ and obtain the second Chern  character only for  four space dimensions \cite{Zhong}.  $Z$ cancels and  we find:
\begin{equation}
 c_{2} =const.\epsilon^{i,j,k,l}Tr[\tilde{G}(\vec{k})\hat{R}_{i}\tilde{G}^{-1}(\vec{k})\tilde{G}(\vec{k})\hat{R}_{j}\tilde{G}^{-1}(\vec{k})\tilde{G}(\vec{k})\hat{R}_{k}\tilde{G}^{-1}(\vec{k})\tilde{G}(\vec{k})\hat{R}_{l}\tilde{G}^{-1}(\vec{k})]
\label{cc}
\end{equation}
The trace operator acts only on the  occupied bands .   The Green's function in the eigen vector basis  representation  replaces the  calculation  with  a multiplicative  \textbf{covariant}  matrix coordinates  $\hat{R}_{i}$. The Chern character   is given by a matrix multiplication.
 The   commutator   $[ \hat{R}_{i},\hat{R}_{j}]=F_{i,j}(\vec{k})$
gives the curvature   $ F_{i,j}(\vec{k})$   in terms of  the connection
$ \mathcal{A}^{(s,s')}_{i}(\vec{k})$. 
\noindent   The second Chern number is given by
\begin{equation}
  C_{2}=\frac{1}{32\pi^2}\int\,d^4k\epsilon^{i,j,k,l}Tr[ F_{i,j}(\vec{k})F_{k,l}(\vec{k})]
\label{top}
\end{equation}
  which  is either  zero (exact form) or non-zero  (non exact form).  Therefore
\begin{equation}
  Tr[ F_{i,j}(\vec{k})F_{k,l}(\vec{k})]\equiv Tr[F^2] = d[K_{3}] 
\label{exact}
\end{equation}
dwhere $K_{3}=Tr[AdA+\frac{2}{3}A^3]$    is the \textbf{Chern-Simons}  three form  \cite{Nakahara} .
 $K_{3}$  can be found with the help of  the  gauge symmetry imposed by  $T^2=-1$ \cite{Nakahara}. 
 The second Chern number in four dimensional space is  given by: $C_{2}=\frac{1}{32\pi^2}\int_{BZ}\,d^{4} k\epsilon^{i,j,k,l} Tr[F_{i,j} F_{k,l}]$  which has a $Z_{2}$  winding number.
 $Tr[F^2]$ is   $closed$ but not $exact$ \cite{Nakahara}.  This means that in some restricted regions of  the Brillouin zone the integral $\int_{S^{4}}Tr[Tr[F^2]$  is  given by  a  Chern-Simons contour integral \cite{Nakahara}:
\begin{equation}
Tr[{F}^2]= d[K_{3}]=d(Tr[\mathcal{A}d\mathcal{A}+\frac{2}{3}\mathcal{A}^3])
\label{chern}
\end{equation}
 This result does not hold  in the entire $B.Z.$.  We can identify two regions which are related by a transition  function (a gauge transformation), a transformation matrix between states  for  the  region $\vec{k}$ which belongs to half of the positive sphere    $(S^{4})_{+}$ and for $\vec{k}$ which belongs to second half, the negative shere   $(S^{4})_{-}$ .   The matrix which transforms between the two regions is  the Pfaffian matrix  $B$ defined in terms of the Kramers pair. 
 We  use the eigenvectors  to compute the  matrix
\noindent
 $B_{ s,s'}(\vec{k})$,  $B_{ s,s'}(\vec{k}) =\langle U_{s}^{(-)}(-\vec{k})|T |U_{s'}^{(-)}(\vec{k})\rangle$ 
\noindent
 the relation between the connections in the different regions $\mathcal{A}(\vec{k})$ and $\mathcal{A}(-\vec{k})$:
\begin{equation}
\mathcal{A}(-\vec{k})= B(\vec{k})\mathcal{A}^{*}(\vec{k})B^{\dagger}(\vec{k}) +i  B(\vec{k})\partial_{k^{i}}B^{\dagger}(\vec{k})
\label{connection}
\end{equation}
and the  curvature transform like :
\noindent
   $ F_{i,j}(-\vec{k})=-B(-\vec{k})  F^*_{i,j}(\vec{k})B(-\vec{k})$.
\noindent
Applying Stokes theorem $ d[K_{3}]$ for the two different regions we obtain a boundary integral over the  difference  of the  Chern-Simons terms defined for each region .The difference between the two Chern-Simons terms  can be understood as a polarization   difference  between  the two regions. The boundary  is a surface perpendicular to   the fourth direction $q$ (physically we can introduce the concept of polarization $P(q)$ which is the difference of the electric flux on the two boundary surfaces perpendicular to   the $q$ direction.)
\begin{eqnarray}
&& C_{2}=\frac{1}{32\pi^2}\int_{BZ}\,d^{4} k\epsilon^{i,j,k,l} Tr[F_{i,j} F_{k,l}]=\int_{  \partial S^{4}}([Tr[\mathcal{A}_{+}d\mathcal{A}_{+}+\frac{2}{3}\mathcal{A}_{+}^3]-Tr[\mathcal{A}_{-}d\mathcal{A}_{-}+\frac{2}{3}\mathcal{A}_{-}^3])\nonumber\\&&
= \frac{1}{24\pi^2}\int\,d^{3}k \epsilon_{i,j,k}Tr[(B(\vec{k})\partial_{i}B^{\dagger}(\vec{k}))(B(\vec{k})\partial_{j}B^{\dagger}(\vec{k}))(B(\vec{k})\partial_{k}B^{\dagger}(\vec{k}))] = \sum_{k^*}N_{k^*}=2P[q]\nonumber\\&&
\end{eqnarray} 
 \noindent We have $2P[q]=\sum_{k^*}N_{k^*}$   ($k^{*}$ are points in the Brillouin Zone  where the Pfaffian matrix vanishes) . Due to the lattice periodicity Bloch theory allows us to define   the polarization as  modulo an integer.  We  recover the result $P=0$ for $\sum_{k^*}N_{k^*}=even$ and  $P=\frac{1}{2}$ for  $\sum_{k^*}N_{k^*}=odd$  \cite{Kane}. $P$  represents the topological invariant for $d=3$ space obtained from the response theory.

\vspace{0.1 in}

\noindent  
 \textbf{3.1.Topological Crystals} 

\vspace{0.1 in}

\noindent 
  This method is  also be applicable to  Topological Crystals where  a mirror  reflection invariant $\eta=(\frac{-i}{\sqrt{2}}(\sigma_{1}+ \sigma_{2})K)$ with the property $\eta^{2}=-1$ replaces the $T^2=-1$ invariant.

\noindent
 Inspired by \cite{Fu} the authors in ref. \cite{Bansil} proposed that $SnTe$ has a  mirror plane perpendicular to the $[110]$ direction.  A band inversion at  the  four  $L$ points  in the Brillouin  zone between $SnTe$ and $PbTe$  can be achieved for the mixed crystal $Pb_{1-x}Sn_{x}Te$. 
The $k\cdot p$ model near an $L$ point \cite{Bansil} is given by:
\begin{equation}
h^{L}=(\sigma_{2}\otimes\tau_{1})k_{1}-(\sigma_{1}\otimes\tau_{1})k_{2}+ (I\otimes\tau_{2})+M(\vec{k})(I\otimes\tau_{3}).
\label{mir}
\end{equation}
where $ M(\vec{k})$ is the inverted mass (has zeros in the Brillouin zone), $\sigma=\pm\frac{1}{2}$ corresponds to the states  with total angular momentum $J=\frac{1}{2}$ , $|j=\frac{1}{2},j_{z}=\pm\frac{1}{2}> $  and $\tau =1,2$  corresponds to the $p$ orbitals  of the cation (Sn or Pb) and anion $Te$.

\noindent 
The mirror invariant $\eta$  with the property   $\eta^2=-1$ can be found  such that  the condition for the polarization is different from the one given  by the time reversal invariant points.

\noindent
 The reflection symmetry from the plane perpendicular to $[1,1,0]$ is given by the transformation $[k_{1},k_{2},k_{3}]\rightarrow [-k_{2},-k_{1},k_{3}]$.  The operator of reflection which acts on the states is given by a rotation of an angle $\pi$ around the axes $[1,1,0]$ and  is accompanied by an inversion trough the origin. 
A simple calculation shows that the mirror operator is given by $M=(\frac{-i}{\sqrt{2}})[(\sigma_{1}+\sigma_{2})\otimes I]$.  As a result the state   $|\phi(\vec{k})>$ is transformed  to $|\phi_{M}(\vec{k}')>$ and the Hamiltonian $h^{L}(\vec{k})$ obeys the symmetry:
\begin{eqnarray}
&&|\phi_{M}(-k_{2},-k_{1},k_{3})>=M|\phi (k_{1},k_{2},k_{3})>\equiv (\frac{-i}{\sqrt{2}})[(\sigma_{1}+\sigma_{2})\otimes I]|\phi (k_{1},k_{2},k_{3})>\nonumber\\&& 
M^{-1}h^{L}(k_{1},k_{2},k_{3})M=h^{L}(-k_{2},-k_{1},k_{3}); \hspace{0.2 in}  M=(\frac{-i}{\sqrt{2}}[(\sigma_{1}+\sigma_{2})\otimes I].\nonumber\\&&
\end {eqnarray}
Next we include the anti unitary  conjugation operator K and define the operator $\eta=MK=(\frac{-i}{\sqrt{2}})[(\sigma_{1}+\sigma_{2})\otimes I]K$ which obeys $ \eta^2=-1$  (this is similar to  the time reversal operator $T=-i\sigma_{2}K$).
We obtain the invariance transformation:
\begin{equation}
\eta^{-1}h^{L}(k_{1},k_{2},k_{3})\eta=(h^{L}(-k_{2},-k_{1},k_{3}))^* .
\label{eta}
\end{equation}
At the     invariance  points  $\vec{k}^{M}$ one obtains the conditions:
 $ h^{L}(-k^{M}_{1},-k^{M}_{2},-k^{M}_{3})= h^{L}(-k^{M}_{2},-k^{M}_{1},k^{M}_{3})$. 
 
\noindent 
For two dimensions we have only two invariant points $[0,0]$ and $[\pi,\pi]$ and for three dimensions we have   four  invariant points $[0,0,0]$, $[0,0,\pi]$, $[\pi,\pi,0]$, $[\pi,\pi,\pi]$.
At this stage we will   follow the strategy proposed in section  $A$. We  identify the pairs of degenerate eigenvalues.
For this case we expect to find a Kramers  pair of eigenfunctions $|V_{s=1}(\vec{k};M(\vec{k})>$ and  $|V_{s=-1}(\vec{k};M(\vec{k}))>$.  When the mass parameter $M(\vec{k}))$ has zeros   we observe that the matrix $\hat{W}_{s,s'}(\vec{k})$ defined by $\hat{W}_{s,s'}(\vec{k})\equiv < V_{s}(-\vec{k};M(-\vec{k}))|\eta|V_{s}(\vec{k};M(\vec{k})>$  is a \textbf{Pfaffian} which vanishes at some $k$'s,  $M(\vec{k})=0$. Due to the zeros of the Pfaffian it is not possible to construct a single eigenfunction for the entire Brillouin zone.  We observe  that the eigenfunction at $-\vec{k}$ is related to the eigenfunction with $ \vec{k}$ in the following way: 
\begin{equation}
|V _{s=1}(-\vec{k};M(-\vec{k})>=\hat{W}^*_{s=1,s'=-1}(\vec{k})|V _{s'=-1}(\vec{k};M(\vec{k})> ;\hspace{0.02 in} |V _{s=-1}(-\vec{k};M(\vec{k})>=\hat{W}^*_{s=-1,s'=1}(\vec{k})|V _{s'=1}(\vec{k};M(\vec{k})>,
\label{Pfafian}
\end{equation}
 where $M(\vec{k})=M(-\vec{k})$.
 For each pair of mirror invariant  states at the   points $[0,0]$ and $[\pi,\pi]$  (for d=2) and  $[0,0,0]$, $[0,0,\pi]$, $[\pi,\pi,0]$, $[\pi,\pi,\pi]$ (for d=3).    The \textbf{Pfaffian} matrix $W$ induces a transformation on the connections,
resulting in the condition for the  Chern-Simons field  polarization $2P[q]=\sum_{k^{M}}N_{k^{M}}$ (with the sum restricted 
to the $\eta$ invariant points).
It results in a different condition for the  polarization,   since $\sum_{k^{M}}N_{k^{M}}$  is different from  the condition $ \sum_{k^{*}}N_{k^{*}}$ for the time reversal case $T=-i\sigma_{2}K$. Therefore we can have a situation where  the polarization is zero according to the time reversal symmetry and non-zero according to the mirror symmetry.

\vspace{0.2 in}

\noindent 
 \textbf{4. Topological invariant for Superconductors}

\vspace{0.2 in}
\noindent
 For superconductors we can not use electromagnetic waves therefore we will use \textbf{sound waves}. In the presence of the  elastic \textbf{crystal} deformation the deformed coordinates gives rise to \textbf{spin-connections} (which are obtained from the derivatives of the metric tensor). Similar to the Electromagnetic  field the spin connection  $\omega_{\mu}^{(a,b)}$ arises from the covariant derivative of the electronic spinors in \textbf{elasticity} \cite{propagating,davidedge},
\begin{equation}
 \partial_{\mu}\Psi^{\alpha}(\vec{r})\rightarrow \bigtriangledown _{\mu}\Psi^{\alpha}(\vec{r})+\frac{1}{4}\omega_{\mu}^{(a,b)}[\gamma^{a},\gamma^{b}]_{\alpha,\beta}\Psi^{\beta}(\vec{r})
\label{elasticity}
\end{equation}
 \cite{davidtop}. The topological invariant being the Pontriagin index  \cite{Nakahara}.
At this point we  use the relation between the second chern character $ch_{2}$ and the Chern-Simons form $K_{3}$, $ch_{2}=d[K_{3}(A,F)]$.
For superconductors we can use the invariance in the entire Brillouine zone obtained by combining the time reversal invariance and the particle-hole symmetry. As a result one can find a unitary  matrix $\Gamma$,  $\Gamma^2=1$ which anti-commutes with the Superconductor Hamiltonian.  One can show that the Hamiltonian can be brought to the form ( due to the Superconducting gap we can use a flat Hamiltonian):
\begin{eqnarray}
&&Q(\vec{k})=\left[\begin{array}{rrr}
                                       0 &q(\vec{k}) \\
q^{\dagger}(\vec{k}) & 0 \\
\end{array}\right] 
\nonumber\\&&
\end{eqnarray}
\noindent From the relation $ch_{2}=d[K_{3}(A,F)]$    we identify the winding number $\nu_{3}=\int_{  S^{4}} d[K_{3}(A,F)]$.
\begin{eqnarray}
&&\nu_{3}=\int_{  \partial S^{4}}(K_{3}[\mathcal{A}_{+},\mathcal{F}_{+}]-K_{3}[\mathcal{A}_{-},\mathcal{F}_{-}])=
 \frac{1}{24\pi^2}\int\,d^{3}k \epsilon_{i,j,k}Tr[(q^{-1}(\vec{k})\partial_{i}q^{\dagger}(\vec{k}))(q(\vec{k})\partial_{j}q^{-1}(\vec{k}))(q(\vec{k})\partial_{k}q^{-1}(\vec{k}))]\nonumber\\&&
\end{eqnarray}
\noindent In one dimensions we have $\nu_{1}=\frac{1}{i2\pi}\int_{B.Z.}dk Tr[q(\vec{k})\partial_{k}q^{-1}(\vec{k})]$.

\noindent
 For two dimensions one identifies the $Z_{2}$ index with the Pfaffian matrix   (the Pfaffian at the time reversal invariant point is equal  to the matrix elements  of the Hamiltonian components $q(\vec{k})$  \cite{Ludwig,Schneider,Ludwigg}):
\begin{equation}
I=\prod_{\vec{\Gamma}}\frac{\mathbf{Pf}\Big[q^{T}[\vec{\Gamma}]\Big]}{\sqrt{det[q[\vec{\Gamma}]]}}
\label{index}
\end{equation} 

\vspace{0.2 in}

\noindent
  \textbf{4.1.Topological invariant for Superconductors from  sound waves  response}

\vspace{0.2 in}


\noindent  
In  the absence of the sound waves  the  $p-wave$  superconductor is given by, 
\begin{eqnarray}
&&H^{(p-wave)}=\frac{1}{2} \int\,d^2r C^{\dagger}(\vec{r},t)\Big[  \tau_{3}\Big(\frac{\hbar^2}{2m} (-i\vec{\partial}_{r})^2 -\mu_{F}(\vec{r})\Big)  -\Delta(\vec{r},t)\Big (\tau^{1}-i\tau^{2}\Big)(\partial_{1}+i\partial_{2})\nonumber\\&&+\Delta^{*}(\vec{r},t) \Big(\tau^{1}+i\tau^{2}\Big)(\partial_{1}-i\partial_{2}) \Big] C(\vec{r},t),
\nonumber\\&&
\end{eqnarray}
$\Delta (\vec{r},t)$ the pairing order field, $\mu_{F}(\vec{r})$ is the space dependent chemical potential  and $\tau^{1}$,$\tau^{2}$ $\tau^{3}$ are the Pauli matrices in the particle-hole space. We assume that in one region    $\mu_{F}(\vec{r})>0$   and in the complimentary   region $\mu_{F}(\vec{r})<0$.  For  $\mu_{F}(\vec{r})>0$ the superconductor is topological and is characterized by the  topological  invariant with the $\mathbf{Chern}$ number $\mathbf{Q}$$\mathbf{\in}$$\mathbf{Z}$ \cite{Alicea}, in the  region  $\mu_{F}(\vec{r})<0$ the superconductor  is non topological. At    the interface  $\mu_{F}(\vec{r})=0$ (between the two regions) the spectrum will contain  bound states,  Majorana zero modes. 
The change of sign of the chemical potential in space gives rise to the Majorana fermions.

\noindent
The presence of sound  waves  deforms the $p-wave$ Hamiltonian 
The sound waves field $\vec{u}(\vec{x},t)$  change  the coordinates     from $\vec{r}\equiv \vec{x}=[x,y]$ to $ \vec{x}+\vec{u}(\vec{x},t)=\vec{X}= \Big[X^{a=1}, X^{a=2}\Big]$  where   $\vec{u}(\vec{x},t)=[u^{(1)}(\vec{x},t),u^{(2)}(\vec{x},t)]$. The strain field are  defined by $E^{i}_{a}=\partial_{a} x^{i}$, $e^{a}_{i}=\partial_{i}X^{a}$ are   related. We have:
 $\sum_{i=1}^{2}e^{a}_{i}E^{i}_{b}=\delta_{a,b}$,   $\sum_{a=1,2}e^{a}_{i}e^{a}_{j}= g_{i,j}$ and $E_{i,b}=g_{i,j}E^{j}_{b}$.
For the time component we have $E^{t}_{i}=e^{t}_{i} \delta_{i,t}$.

\noindent 
The derivatives transform like vectors, $\partial_{a}=\partial_{a} u^{1}(\vec{x},t)\partial_{1}+ \partial_{a} u^{2}(\vec{x},t)\partial_{2}$.  For   the sound waves   $\vec{u}(\vec{r},t)$ we have:   $e^{1}_{i}=\delta_{1,i}-\partial_{i}u^{1}(\vec{x},t)$, $e^{2}_{i}=\delta_{2,i}-\partial_{i}u^{2}(\vec{x},t)$,   $e^{a}_{t}=-\partial_{t}u^{a}(\vec{x},t)$ for $ a=1,2$.
The integration area element $d^2x$ in   Eq.$(2)$ is  multiplied by the Jacobian $J=  \Big[e^{1}_{1}e^{2}_{1}- e^{1}_{2}e^{2}_{1}\Big]\equiv Det[e^{a}_{i}]$.   The deformed $p-wave$  Hamiltonian $H^{(deformed-p-wave)}$ takes the form:
\begin{eqnarray}
&&H^{(deformed-p-wave)}= \frac{1}{2}\int\,d^2x Det[e^{a}_{i}] C^{\dagger}(\vec{r},t)\Big[  \tau^{3}\Big(\frac{-\hbar^2}{2m} (\sum_{a=1}^{2}E^{i}_{a}E^{j}_{a} \nabla_{i}  \nabla_{j}) -\mu_{F}(\vec{x})\Big)  -\Delta(\vec{x},t)\Big (\tau^{1}-i\tau^{2}\Big)(E^{i}_{1}\nabla_{i}+iE^{i}_{2}\nabla_{i})\nonumber\\&&+\Delta^{*}(\vec{x},t) \Big(\tau^{1}+i\tau^{2}\Big)(E^{i}_{1}\nabla_{i}-iE^{i}_{2}\nabla_{i}) \Big] C(\vec{x},t)\nonumber\\&&
\end{eqnarray}
\noindent
 $\mathbf{\nabla}_{i}$ is  the covariant derivative  given in terms of the spin connection:
\noindent
 $\mathbf{\omega}_{i}^{a,b}[\tau^{a},\tau^{b}]$, $\mathbf{\nabla}_{i}\equiv \partial_{i}+\frac{1}{8}\mathbf{\omega}_{i}^{a,b}[\tau^{a},\tau^{b}]$.  The spin connection has been derived  in terms of  $E^{i}_{a}$ and $e^{a}_{i}$. The  spin connection  is determined  from the zero torsion condition, $\nabla_{i}e^{a}_{j}- \nabla_{j}e^{a}_{i}=0 $. We have,
\begin{equation}
\omega^{a}_{i}=\epsilon^{a,b,c}E^{j}_{c}(\partial_{i}E_{j,b}-
\partial_{j}E_{i,b})-\frac{1}{2}\epsilon^{b,c,d}(E^{j}_{c} E^{k}_{c}\partial_{k} E_{j,d})e^{a}_{i}
\label{notorsion}
\end{equation}
Where $E_{i,b}=g_{i,j}E^{j}_{b}$.
 
Once the spin connection is know we can perform the path integral over the Dirac' fermion. As for the Yang-Mills theory  the  fermion integration in $2+1$ dimensions generate  a  non-Abelian Chern-Simons  term.  
We perform the path integral   integration  tor the fermion field  $C^{\dagger}(\vec{r},t)$  and  $C(\vec{r},t)$  and obtain the effective sound action $S^{top-sound}$  \cite{propagating}.    The topological term is given by, 
\begin{equation}
S^{top-sound}=\frac{c}{96\pi}\int\,dt\int\,d ^2r\Big[\epsilon^{i,j,k}\omega_{i,a}(\partial_{j}\omega^{a}_{k}-\partial_{k}\omega^{a}_{j})+\frac{2}{3}\epsilon^{a,b,c}\omega_{i,a}\omega_{i,b}\omega_{i,c}\Big]
\label{topolo}
\end{equation}
\noindent 
 where $\omega_{i,c}=g_{i,j}\omega^{j}_{c}$.
$\mathbf{c}$ counts the number  of the Majorana  edge  modes. For the Topological Superconductor we  have  $\mu_{F}>0$ and $\mathbf{c}$ is non  zero. For this case Majorana modes are on the edge of the sample. As a result the effective  sound action $S^{top-sound}$ allows to identify the Topological Superconductor.  It is important to mention that the  connections $ \omega^{a}_{i}=\omega^{a}_{i}(t)$ are function of the driving force which excites the solid.  Due  to the high order derivatives it is difficult to  observe such   terms in  the laboratory. For this reason we will look for an alternative way to identify the Topological Superconductor.
This result is similar to the  gravitational Chern-Simons term.

\vspace{0.2 in}

\noindent 
 \textbf{5. Equation of motion in the B.Z. for non commuting coordinates }

\vspace{0.2 in}

\noindent 
The methodology of a "curved" space induced by  the spin orbit coupling in the B.Z. was  described  in  terms  of the connection  $ A^{\alpha,\beta}_{a}(\vec{k})\equiv  \langle U_{\alpha}(\vec{k})|i\partial_{a}|  U_{\beta}(\vec{k})\rangle=\int\,d^{d}x U^{*}_{\alpha,k}(x)( x^{a}) U_{\beta,k}(x)$ and the  \textbf{curvature} $[X_{a},X_{b}]=F_{a,b}$  for the coordinates. This allows to introduce   the first and second \textbf{chern character}   $ch_{1}= \frac{i}{2\pi}\mathbf{Tr}\Big( F_{a,b}(\vec{k})\Big)$,  $ch_{2}= \frac{-1}{32 \pi^2} \epsilon^{a,b,c,d}\mathbf{Tr}\Big( F_{a,b}(\vec{k}) F_{c,d}(\vec{k})\Big)$  which are   given in terms of the \textbf{covariant}  coordinate  (or polarization)  $X_{a}=x_{a}+A_{a}(\vec{k})$ ( $ x_{a}$ is the coordinate in the momentum space   $ x_{a}=i\partial_{k^{a}}$) . 
 The approach is  is based the  non-commuting covariant coordinates , $F_{a,b}=[X_{a},X_{b}] $.  
For example: in the eigenvalue representation the Hamiltonian is given by: $\mathbf{h}(\vec{k})=\sum_{\lambda}E_{\lambda}(\vec{k})|U_{\lambda}(\vec{k})><U_{\lambda}(\vec{k})|$, the presence of a scalar potential $\mathcal{V}(\vec{r})$ is replaced by $\mathbf{V}(\vec{R})$  where  $\vec{R}$ is the matrix covariant coordinate. The Heisenberg equation of motion are given by:
\begin{eqnarray}
&&\frac{d k^{a}}{dt}=\frac{-1}{\hbar}\partial_{X_{a}}\mathbf{V}(\vec{X}) \nonumber\\&& \frac{d X_{a}}{dt}=\frac{1}{\hbar}\partial_{k^{a}} \mathbf{h} (\vec{k})+\frac{1}{\hbar}\epsilon^{a,b,c}F_{b,c}\partial_{X_{a}}\mathbf{V}(\vec{X})\nonumber\\&&
\end{eqnarray}
When the real space is curved, due to dislocations or magnetic fields the momentum operator $k^{a}$ is replaced by a covariant momentum matrix $K^{a}=k^{a}-\frac{i}{4}Log[g]$ where  $g=\sqrt{ Det g_{\vec{r},t}}$ is the metric tensor  \cite{davidSpinorbit}. As a result, the real  space curvature is given by,
\begin{equation}
\Omega_{a,b}=[K^{a},K^{b}]
\label{real}
\end{equation}
The equation of motion for the covariant  momentum is accordingly  modified:
\begin{eqnarray}
&&\frac{d K^{a}}{dt}=\frac{-1}{\hbar}\partial_{X_{a}}\mathbf{V}(\vec{X}) +\frac{1}{\hbar}\epsilon^{a,b,c}\Omega_{b,c}\partial_{K^{a}}\mathbf{h}(\vec{K})\nonumber\\&&  \frac{d X_{a}}{dt}=\frac{1}{\hbar}\partial_{K^{a} }\mathbf{ h}(\vec{K})+\frac{1}{\hbar}\epsilon^{a,b,c}F_{b,c}\partial_{X_{a}}\mathbf{V}(\vec{X}) \nonumber\\&&
\end{eqnarray} 
The basic ingredients are the commutators in the $k$ space  $\Omega_{a,b}=[K^{a},K^{b}]$ for the momentum and $F_{a,b}=[X_{a},X_{b}] $ for the coordinates.

\vspace{0.2 in}

\noindent 
 \textbf{6. Topological invariant in  two space dimensions derived from  topological  invariant in four space dimensions.}

\vspace{0.2 in}

\noindent
The  topological response for time reversal invariant systems in one and two space dimensions is not entirely clear .  In three space dimensions we can use the   Chern-Simons form $ K_{3}(A,F)]$     to relate the the second Chern number $ C_{2}$ in  four space dimensions to tree dimensions using the relation  $ch_{2}=d[K_{3}(A,F)]$.
In four dimensional  momentum space the second Chern number $ C_{2}$ is   given by an \textbf{ index  operator}.
In analogy with the index operator for the Dirac equation I introduce the index operator in the momentum space $Ind.\Big[i\mathbb{R}_{4}\Big]$:
\begin{equation}
 Ind.\Big[i\mathbb{R}_{4}\Big] =  Tr\Big[\gamma^{5 } e^{-\epsilon \int\,d^4k \bar{ C}(\vec{k})\Big( i\mathbb{R}_{4}(\vec{k})\Big)^2 C(\vec{k})}\Big] |_{\epsilon \rightarrow 0 }\longrightarrow  C_{2}
\label{Indexmiatrix}
\end{equation} 
\noindent 
The operator 
$i\mathbb{ R}$ is defined in terms of the non-Abelian  spin connection :
\begin{eqnarray}
 &&A^{\alpha,\beta}_{a}(\vec{k})\equiv  \langle U_{\alpha}(\vec{k})|i\partial_{a}|  U_{\beta}(\vec{k})\rangle\nonumber\\&&
i\mathbb{ R}_{4}(\vec{k})= i\sum_{a=1,2,3,4}(\gamma_{a}(x^{a}+A_{a}(\vec{k}))\equiv i\gamma_{a}X^{a}(\vec{k})\nonumber\\&&
 \gamma^{5 }=\left[\begin{array}{rrr}
                                       -1 & 0 \\
0 & 1 \\
\end{array}\right] \nonumber\\&&
\end{eqnarray}
$ \gamma^{5 }$ separates the conduction band from the valence band.

\noindent  In order to show that the index operator in four space dimensions  is related to the  index  operator in d=2 space  dimensions, we introduce the transformation :
\begin{equation} 
\mathcal{A}^{g(\theta)}(\vec{k})=g^{-1}(\theta,\vec{k})\Big(\mathcal{A}(\vec{k})+d\Big)g(\theta,\vec{k}) 
\label{conections}
\end{equation}
where $\theta$ parametrizes the $S^{1}$ circle (see  figure $13.4$  page $521$  \cite{Nakahara}). 
Next we construct a family of gauge fields: 
\begin{equation}
\mathcal{A}^{t,\theta}\equiv t \mathcal{A}^{g(\theta)};\hspace{0.1 in} 0\leq t \leq 1
\label{t}
\end{equation}

\noindent  The parameters $(t,\theta)$  form a disc $D^2$ with $\partial D^2=S^{1}$ .
We construct from$ D^2\times S^2$ a manifold  $S^2\times S^2$ . We will call the patch $(t,\theta) $ the northern hemisphere $U_{N}$ and  $(s,\theta)$  the suthern hemisphere $ U_{s}$ and the equator $S^{1}$ of $S^{2}$ corresponds to $t=s=1$ (see the  figure with the two half sphere,  figure $13.4$  page $521$  in \cite{Nakahara}).
The gauge potential in $2+2$ dimensions can be written as:
\begin{eqnarray}
&&\mathbb{A}_{N}(t,\theta,\vec{k})=\Big[0,0,\mathcal{A}^{t,\theta}(\vec{k})+g^{-1}(\theta,\vec{k})d_{\theta}g(\theta,\vec{k})\Big]\nonumber\\&&
\mathbb{A}_{N}(t,\theta,\vec{k})=\Big[0,0,\mathcal{A}(\vec{k})\Big]\nonumber\\&&
\end{eqnarray}
\noindent   $\mathcal{A}(\vec{k})$ is  the non-Abelian spin connection in two space dimensions. On the equator $t=s=1$ we have :
\begin{equation}
\mathbb{A}_{N}=g^{-1}\Big(\mathbb{A}_{s}+d+d_{\theta}+d_{t})g
\label{spin}
\end{equation}
 where $d$ is the \textbf{exterior derivative} \cite{Nakahara} in two dimensions, $d_{\theta}$ is the external derivative on $S^{1}$ and $ d_{t}g=0$. $\mathbb{A} =(\mathbb{A}_{N},\mathbb{A}_{S})$  is the spin connection in $2+2$ dimensions.
We use the relation,
\begin{equation}
 det\Big[i\hat{D}(\mathcal{A}^{g(\theta)})(\vec{k})\Big]\equiv\Big[\det(i\mathbb{R}(\mathcal{A}(\vec{k}))\Big]^{\frac{1}{2}}e^{iw(\mathcal{A},\theta)}
\label{det}
\end{equation}
 $|det[i\hat{D}]|$ is gauge invariant and only  $det[i\hat{D}]$ might have  anomalous behavior . On the boundary disc $\partial D^2=S^{1}$  the phase $e^{iw(\mathcal{A},\theta)}$ defines the maping  $\partial D^2\rightarrow S^{1}$
\begin{equation}
 i\hat{D}(A)= i\sum_{a=1,2,3,4}(\gamma_{a}(x^{a}+iA_{a}(\vec{k})P_{+}) ,\hspace{0.1 in} P_{+}=\frac{1}{2}(1+\gamma^{5 })
\label{anomalous}
\end{equation}
 $|det [i\hat{D}(A)]|$ is gauge ivariant and only the phase  $ w(\mathcal{A},\theta)$ is anomalous and gives    the winding number $\nu_{1}$, on the disc $D^2$ there are points at which  $det [i\hat{D}(A)]$   vanishes.
\noindent  The  index  $Ind.\Big[i\mathbb{R}_{2+2}\Big]$ is given by,
\begin{equation}
Ind.\Big[i\mathbb{R}_{2+2}\Big]=\int_{S^2\times S^2} c_{2}(\mathbb{F})=\nu_{1}  
\label{nu}
\end{equation}
 where  the curvature $\mathbb{F}$ is given by,
\begin{equation}
 \mathbb{F}=(d+d_{\theta}+d_{t})\mathbb{A}+\mathbb{A}^2
\label{F}
\end{equation}
 $\nu_{1}$ is the winding number   which  is  is even  or odd and corresponds to the index $Z_{2}$ introduced earlier.

\noindent 
Folowing the procedure used before  which relates the Chern character  to the Chern-Simons form  $c_{2} = d[K_{3}(A,F)]$  we find:
\begin{equation}
\int_{S^2\times S^2} c_{2}(\mathbb{F})=\int_{D^2\times S^2}
 c_{2}(\mathbb{F}_{N})+\int_{D^2\times S^2}
 c_{2}(\mathbb{F}_{S})=\int_{S^{1}\times S^2}\Big[K_{2+1}(\mathbb{A}_{N},\mathbb{F}_{N})|_{t=1}-K_{2+1}(\mathbb{A}_{S},\mathbb{F}_{S})|_{s=1}\Big]
\label{procedure}
\end{equation}
Since $\int_{S^{1}\times S^2} K_{2+1}(\mathbb{A}_{S},\mathbb{F}_{S})=0$
we find:
\begin{equation}
Ind.\Big[i\mathbb{R}_{2+2}\Big]=\int_{S^{1}\times S^2}K_{2+1}(\mathbb{A}_{N},\mathbb{F}_{N})=\int_{S^{1}\times S^2}K_{2+1}(\mathcal{A}^{g(\theta)}+g^{-1}d_{\theta}g,\mathcal{F}^{g(\theta)})=-(\frac{1}{2\pi})^2 Tr[W d\mathcal{A}]
\label{result}
\end{equation}
\noindent
 where $ W\equiv g^{-1}(\theta)d_{\theta}g(\theta)$. For $\theta=0$  we find  $W=1$ and   $Ind\Big[i\mathbb{R}_{2+2}\Big]= \nu_{1}$   ($\nu_{1}$ is  the winding  number )  which is identical to the $Z_{2}$ index introduced by \cite{Kane}.

\noindent 
The procedure presented here  allow a direct construction of the $Z_{2}$ invariant in two dimensions as an emergent object from  four dimensions and therefore is related to the electromagnetic or sound wave response defined in four space dimensions. Contrary to early procedures which used dimensional reduction the  procedure proposed  is based on deforming  the spin connection $\mathcal{A}$ to a family of higher dimensions gauge potentials $\mathcal{A}^{g(\theta)}$.


\vspace{0.2 in}

\noindent 
\textbf{7. Chiral $p-wave$ wire wave coupled to two metallic rings pierced by flux- Detection of the Majorana Fermions  by measuring the persistent current}

\vspace{0.2 in}
\noindent 
We consider a situation where two metallic rings are attached to the   two ends of the  $ p-wave$  (which has  two zero modes at the of the wire $\gamma_{1}$ and $\gamma_ {2}$) (see \cite{rings,davidMajorana}).
\noindent  We consider the special case where $ L=Nl$ ($L$ is the wire length and  $ l $   lenght of each  ring )
The flux in ring one is  $ f_{1}$ and in   ring  two is  $ f_{2}$. Using   periodic boundary  conditions  $ C_{1}(-l)=C_{1}(0)$   and  $ C_{2}(-l)=C_{2}(0)$  we perform a gauge transformation .    Due to  the flux we   obtain  twisted  boundary conditions, $\hat{C}_{1}(x)=e^{i\frac{2\pi x}{l}f_{1}}C_{1}(x) $, $\hat{C}_{2}(x)=e^{i\frac{2\pi x}{l}f_{2}}C_{2}(x) $ which allows to   find the spectrum of  the two uncoupled rings  as a function of the momentum $ \frac{2\pi}{l}n$, $ n=0,\pm1,\pm 2,...$.
\begin{equation}
H_{0}=\sum_{n}\frac{\hat{\gamma}}{2}\hat{C}_{1}^{\dagger}(n)(n+f_{1})^{2}\hat{C}_{1}(n) +\sum_{n}\frac{\hat{\gamma}}{2}\hat{C}_{2}^{\dagger}(n)(n+f_{2})^{2}\hat{C}_{2}(n);\hspace{0.01in}H^{Maorana}=\epsilon_{0}\zeta^{\dagger}\zeta 
\label{esp}
\end{equation}

\noindent
$H^{Maorana}$ is the Hamiltonian of the $p-wave$ wire restricted only to the Majorana modes. Expressed in terms of the fermion fields $\gamma_{1}=\frac{1}{\sqrt{2}}\Big[\zeta^{\dagger}+\zeta\Big]$ ; $\gamma_{2}=\frac{1}{i \sqrt{2}}\Big[\zeta^{\dagger}-\zeta\Big]$

\noindent
The coupling between the wire and the ring are  given by :
\noindent
$H_{t}=\frac{g}{\sqrt{2}}\sum_{n}\Big[(\hat{C}_{1}^{\dagger}(n) (\zeta+\zeta^{\dagger})  +  (\zeta+\zeta^{\dagger}) \hat{C}_{1}(n))  \Big]+(-i)\frac{g}{\sqrt{2}}\sum_{n}\Big[(\hat{C}_{2}^{\dagger}(n) (\zeta-\zeta^{\dagger})  +  (\zeta-\zeta^{\dagger}) \hat{C}_{2}(n))  \Big]$

\noindent 
We will integrate the rings degree of freedom and obtain an effective Majorana  impurity Hamiltonian.
\begin{equation}
 H_{eff} =\hat{\zeta}^{*T}(\omega)  \Big[ M(\omega;f_{1},f_{2})\Big]\hat{\zeta}(\omega) ;  \hat{\zeta}^{*T}(\omega) =\Big[\zeta^{\dagger}(\omega),\zeta,(\omega),\zeta^{\dagger}(-\omega),\zeta(-\omega)
\Big]
\label{impurity}
\end{equation}
\noindent
$M(\omega)$ is a $ 4 \times 4$ matrix which depends on $\Delta$ and $\delta$  :
$M_{1,1}(\omega)=  \omega-\epsilon_{0}-\Delta $, $M_{1,2}(\omega)=M_{1,3}(\omega)=0$,$ M_{1,4}(\omega)=-\delta$;
$M_{2,1}(\omega)=0 $, $M_{2,2}(\omega)=-(\omega+\epsilon_{0})+\Delta$, $M_{2,3}(\omega)=\delta$,  $M_{2,4}(\omega)=0$;
$M_{3,1}(\omega)=0$ ,$ M_{3,2}(\omega)=\delta$, $ M_{3,3}(\omega)=  -(\omega+\epsilon_{0})+\Delta$, $M_{3,4}(\omega)=0$;
$M_{4,1}(\omega)=\delta $,$M_{4,2}(\omega)=M_{4,3}(\omega)=0$, $M_{4,4}(\omega)= (\omega+\epsilon_{0})-\Delta$;
$\Delta=\frac{g^2}{8}(\Delta^{(1)}+\Delta^{(2)})$, $\delta=\frac{g^2}{8}(\Delta^{(1)}+\Delta^{(2)})$,
$\Delta^{(1)}=\sum_{n}\Big[\frac{\omega-E_{1}(n) +ix}{(\omega-E_{1}(n))^2+x^2}\Big]$ , $\Delta^{(2)}=\sum_{n}\Big[\frac{\omega-E_{2}(n) +ix}{(\omega-E_{2}(n))^2+x^2}\Big]$,
$E_{1}(n)=\frac{\kappa}{2}(n+f_{1})^{2}-\mu$;
$E_{2}(n)=\frac{\kappa}{2}(n+f_{2})^{2}-\mu$.
\noindent
We integrate the Majorana Fermion and obtain the exact partition function:
$\mathbf{Z}(f_{1},f_{2})=Z(g=0;f_{1},f_{2})\cdot det[M(\omega;f_{1},f_{2})]$

\noindent
 where $ Z(g=0)$ is the partition function of two uncoupled rings.
Therefore the current is given by $I_{1}(f_{1},f_{2})=\Big[ \frac{ d Log ( \mathbf{Z}(f_{1},f_{2}))}{df_{1}}\Big]$     ; $I_{2}(f_{1},f_{2})= \Big[ \frac{ d Log ( \mathbf{Z}(f_{1},f_{2}))}{df_{2}}\Big]$.
Due to the multiplicative form of the partition function the current is a sum of two parts  $I_{i}(f_{i}; g=0)$ (i=1,2)  and a second   part which  is determined  by  the matrix  $ M(\omega;f_{1},f_{2}) $   and  is given by    $I_{i}(f_{1},f_{2})$ , $i=1,2$ .
The current in each ring  is given by,  $I_{1} (f_{1},f_{2})=I_{1}(f_{1}; g=0)  + I_{1}(f_{1},f_{2})$ (first ring) $I_{2} (f_{1},f_{2})=I_{2}(f_{2}; g=0)  + I_{2}(f_{1},f_{2})$(second ring).

\noindent 
We investigate  the case of  equal fluxes, $ f_{1}=f_{2}=f$.Due to the fact that $ L=Nl$ the hoping matrix elements are real.  In particular the Majorana energy $\epsilon_{0}$ couple like a regular impurity to a set of states determined by the two rings . Effectively the integration of the electrons in the rings  renormalizes  the energy $ \epsilon_{0}$to   $\ epsilon_{eff.}(\epsilon_{0},f)=\epsilon_{0}+\Sigma(f)$   where $\Sigma(f) $ is the shift in energy caused by the energy in the ring $E(n,f)$.

\begin{figure}
\begin{center}
\includegraphics[width=2.5 in ]{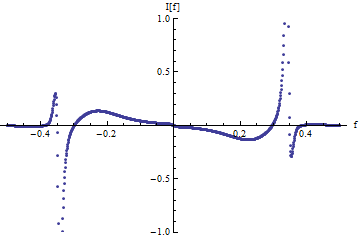}
\end{center}
\caption{The shift of the  persistent current  for the Majorana energy $\epsilon_{0}=0.1$ .}
\end{figure}
The current for a single ring is computed  in the absence of $ p-wave$ wire .  The current was computed for a fixed chemical potential, this explains the jump of the current  at $ f=\pm0.4$.
\noindent
In the presence of the Majorana energy    $\epsilon_{0}=0.1$  in  figure $(1)$   we obtain a shift    in the  persistent current  with respect  the single ring .For the Majorana energy    $\epsilon_{0}=0.01$  the shift in the persistent current is reduced (the jump of the current  $ f=\pm0.4$ originate from the fixed chemical potential  as seen for the single ring).  When the Majorana energy goes to zero the only contribution to the persistent  current comes  from the single  ring . This dependence  is due to the ground state energy which   changes with the flux.
In the presence of a Majorana Fermion we have  in addition to  the persistent  current which  comes  from the perfect   wire  contributions  like  $epsilon_{eff.}(\epsilon_{0},f)=\epsilon_{0}+\Sigma(f)$     a contribution from  the matrix   $M(\omega)$  which is a function of the Majorana energy determines and  determines  shift of the persistent current.

\noindent
The persistent current is measured by the scanning  of the \textbf{SQUID} \cite{Moler}  which measures the  change in the magnetization.  The magnetization is proportional  to the persistent current.  One subtracts   from  the persistent  current  contribution  from the single ring    and find  the dependence on the Majorana energy. 
 In the past only measurements of few rings was possible .
In recent years some of the experimental groups have claimed   to measure the persistent current  in a single ring.  Some of the complications  are  related to the fact that the measurements have been   done in diffusive limit (with 10-20 rings)  and not in the ballistic limit which we assumed in our calculation. 
The effect of the diffusive limit ca be study by coupling the rings to a  noise bath.

\vspace{0.2 in}

\noindent 
\textbf{8.  Surface Physics-Photoemission}

\vspace{0.2 in}

\noindent
One of the ways to study  the topology of the $T.I.$ is to investigate the zero modes.  The surface  states \textbf{zero modes} (edge states )   for  a  $TI$  have an odd number of   chiral modes  $ N_{right}-N_{left}= odd$. The surface mode for the $3d$  $T.I.$  is given by one chiral mode, which is   the Weyl Hamiltonian \cite{davidT}. The  conduction electrons have  a fixed chirality  on the surface of the  $T.I.$.  One observes   topological surface excitations which are characterized by the integrated Fermi surface  Berry curvature of $\pi$.  As a result,  localization is prohibited  \cite{davidT}.  The  optical conductivity,  the  Raman spectrum,    the polarization  of  the  photoeletrons  and the  photoconductivity   can reveal    the topology  of the surface  states and spin texture.
Recently  the use of an innovative  spectrometer with a high laser-based light source  has shown that the spin polarization of the  photoelectrons  emitted from the surface of the  $ Bi_{2}Se_{3}$  Topological Insulator ($T.I.$)  can be manipulated  through the  laser light  polarization \cite{Nature}. One finds that the photoelectrons  polarization  is completely  different  from the  initial states  and is controlled by the  photon polarization.
The interpretation of these  experiments are based on the assumption that  the surface of the $T.I.$ is flat.
We will   to compute  the     detection  of photoelectrons   and photoconductivity  for  an arbitrary   a crystal-face  boundary. 
We will   introduce a  model  for  \textbf{photoemission} for for an arbitrary surface.

\vspace{0.1 in}

 \noindent 
\textbf{ 8.1. Surface states  for   an arbitrary crystal-face boundary}

\vspace{0.1 in}
 \noindent 
 Photoemmision, photoconductivity, optical conductivity   and scanning tunneling microscopy      are sensitive  to the nature    of the surface states. The topology of the surface states is affected by the physical boundary. For an arbitrary surface we need to solve the problem using a  curved coordinate   bases which rotates from point to point.  For this type of problems  the  \textbf{non-coordinate bases} $\partial_{a}$, $a=1,2,3$ introduced by Cartan \cite{Nakahara,Nieh} is  related to   the Cartesian directions $\partial_{\mu}$ , $\mu=x,y,z$.
For example a point on a surface  is given by the vector  $\vec{r}=[x(u_{1},u_{2}),y(u_{1},u_{2}),z(u_{1},u_{2})]$  and   by   the  coordinates  on a surface   $ (u_{a=1},u_{b=2})$  with  the  normal   to the surface  $ E_{N}$.  The  two sets of coordinates  are related  throught  the matrix  $E^{\mu}_{a}=  \frac{\partial r^{\mu}}{\partial u_{a}}$, $\mu=x,y,z$ and $a=1,2$.  The normal to the surface  is given by  $ E_{N}=\frac{\frac{\partial \vec{r}}{\partial u_{1}} \times \frac{\partial \vec{r}}{\partial u_{2}}}{||\frac{\partial \vec{r}}{\partial u_{1}} \times \frac{\partial \vec{r}}{\partial u_{2}}||}$.
This set of transformations  allows to replace $\sigma^{a}\partial_{a}$  by the covariant derivative $\nabla_{a}= \partial_{a}+ \frac{1}{8}\Gamma^{b,c}_{a}[\sigma_{b},\sigma_{c}]$  (which  depends on the connection defined below) ; 
$ \sigma^{a}\nabla_{a}=\sigma^a E^{\mu}_{a}\Big[\partial_{\mu}+\frac{1}{8}\Gamma^{a,b}_{c}e^{c}_{\mu}[\sigma_{a},\sigma_{b}]\Big]$, $E^{\mu}_{a}$ and $ e^{c}_{\mu}$  are the transformation  and the inverse transformation matrix and  $ \Gamma^{a,b}_{c}$ are the \textbf{connection one form} matrix (see \cite{Nakahara}  page 285).

\noindent
 We  will consider first the  Weyl Hamiltonian for  cylindrical coordinates (this problems has been considered in the literature   \cite{Ashvin,hanaguri, Franzaxion}and  here we will introduce a different method in order to deal with the  arbitrary crystal face boundary.

\noindent  
 \textbf{The Weyl Hamiltonian in Cartesian coordinates  is} :$ H^{2d}=(-i)[\sigma_{x}\partial_{y}- \sigma_{y}\partial_{x}]$.

\noindent 
 We put the Hamiltonian on a cylinder and take the axes $x=x^{1}$;  $y=x^{2}= r \sin[\phi]$  and  $z=x^{3}= r \cos[\phi]$;
$\vec{r}=\Big[x, r \sin[\phi],r \cos[\phi]\Big]$.
 We propose to study this problem using  the \textbf{non-coordinate basis} given by Cartan  \cite{Nakahara}: $u_{ a=1}=x$ ,$ u_{ a=2}=\phi$ and $ r$ is the coordinate in the normal direction $E_{N}$; the \textbf{derivatives} are 
$\hat{\partial}_{a=1}=\partial_{x}$;
$\hat{\partial}_{a=2}=\frac{1}{r}\partial_{\phi}$;$\hat{\partial}_{N=3}=\partial_{r}$ and the \textbf{differentials (one form)}  are given by:
$\theta^{1}=dx$;  $\theta^{2}=r d\phi$;  $\theta^{3}=dr$.
The coordinate basis is not fixed,  therefore \textbf{ connection}   $\Gamma^{a}_{a,b}$  will be generated.
The  connection $\Gamma^{a}_{a,b}$  are  determined  from  the Cartan's structure equation   for the \textbf{Torsion} $T^a$ : $d \theta^{a}+\omega^{a}_{b}\wedge \theta^{b}=T^{a}$, $a=1,2,3$. The connection $ \omega^{a}_{b} $ is expanded in terms of the differential $\Gamma^{a}_{b,c}\theta^{c}$ one form with the help of  the matrix transformation  $e^{c}_{\mu}$:
\noindent
  $  e^{c}_{\mu}E^{\mu}_{a}=\delta^{a}_{c}$ ;$E^{\mu}_{a}\equiv\partial_{a} 
\vec{r}=
\Big[\partial_{a}x,\partial_{a}r \sin[\phi],\partial_{a}r \cos[\phi]\Big] $ ; 
\noindent
$E^{\mu}_{1}=\Big[1,0,0\Big]$;
$E^{\mu}_{2}=\Big[0,r \cos[\phi],-r\sin[\phi]\Big]$;$E^{\mu}_{2}=\Big[0,\sin[\phi], \cos[\phi],\Big]$

\noindent 
 From the transformation we obtain :
$\omega^{a}_{b}$  and $\omega^{a,b}_{\mu}=\Gamma^{a,b}_{c}e^{c}_{\mu};
\hspace{0.1 in}\omega^{a,b}_{\mu}=-\omega^{b,a}_{\mu}$

\noindent
 From the torsion condition    $\textbf{Torsion}=T^{a} =0$   we find:
$d \theta^{a}+\omega^{a}_{b}\wedge \theta^{b}=T^{a}=0$,   ( $\wedge $ is the wedge product \cite{Nakahara}) $a=1,2,3$.  We obtain the      equation:
$d \theta^{a}=-\Gamma^{a}_{b,c}\theta^{c}\wedge\theta^{b}$
;$ \Gamma^{2}_{2,3}=-\Gamma^{2}_{3,2}=\frac{1}{r}$.

\noindent
 \textbf{the Weyl Hamiltonian in the cylindrical basis} is given by:
\noindent
$H^{cyl.}=    \sum_{a=1,2,3}(-i)\sigma^{a} \nabla_{a}\equiv  \sum_{a=1,2,3}(-i)\sigma^{a}\Big[\hat{\partial}_{a}+\frac{1}{8}\Gamma^{a}_{a,b}[\sigma_{a},\sigma_{b}]\Big]=
(-i)[\sigma^{1}\partial_{x}+\sigma^{2}\frac{1}{L}\partial_{\phi} - \frac{1}{2 r}\sigma^{3}]$

\noindent 
The eigenvalue equation $H^{cyl.}\psi(x,\phi)=E \psi(x,\phi)$ has  real solutions for  boundary conditions $\psi(x,\phi+2\pi)=-\psi(x,\phi)$.  We find: $E=\pm\sqrt{k^2_{x}+\frac{l(l+1)}{r^2}}$;
$\psi^{\pm}(x,\phi)=e^{i k_{x}x +(l+\frac{1}{2})\phi}\Big[1,e^{-i[\pm\chi(k_{x},l)-\alpha(k_{x},l)]}\Big]^{T}$,  $k_{x}=\frac{2\pi}{L}n$ $,n=0,\pm1,\pm2,..$;     $ l= 0,\pm1,\pm2,..$ ;$ tan[\chi(k_{x},l)]=\frac{l+\frac{1}{2}}{k_{x}r}$;  $tan[\alpha(k_{x},l)]=\frac{1}{2r\sqrt{k^2_{x}+\frac{l(l+1)}{r^2}}}$

\vspace{0.1 in}

\noindent
\textbf{8.2. The Hamiltonian  for  crystal face   with a  cylindrical  boundary}

\vspace{0.1 in} 

\noindent
In order to understand why the curved coordinates emerge,  we will consider   a three dimensional  $T.I.$    with the mass dependent gap  is in the $z$ direction. Such a $T.I.$  has a surface boundary   perpendicular to the  axes $z$ with crystal-face $[x,y]$ plane  localized at $z=L$.
To simplify the discussion,  we consider a situation where the crystal-face   is  cylindrical with the  cylindrical  axis  in the $ x$ direction and  length $L_{x}>L$.  As a result, any  point on the surface of the cylinder is given  by the  set of coordinates   $\vec{r}=[x,L\sin(\phi),L\cos(\phi)]$.
\noindent
The four-band model for the three dimensional $T.I.$, $Bi_{2}Se_{3}$  \cite{Kane}   of size $L_{x}\times L\times L$ .  
is given by $ H=H_{\bot}+H_{\|}$ where  $\tau$ are the Pauli matrix for the orbitals and  $\sigma$   represents the spin:

\noindent
$H_{\bot}=\tau_{3}(-m_{0}(z)-m_{2}\partial^2_{z}) +i\tau_{2}\partial_{z}$;
$H_{\|}=(-i)\tau_{1}(\sigma_{2}\partial_{y}- \sigma_{1}\partial_{x})-m_{\|}\tau_{3}(\partial^2_{x}+\partial^2_{y})$

\noindent
 The mass is a function  of $z$,  $ m_{0}(z)=|m_{o}|F[-z+L]+(-M)F[z-L]$,  where $F[z]$ is the step function.  We have for $ z>L$, $ M\rightarrow \infty$, therefore  we obtain a  
 zero mode on the surface,  given by the  two dimensional Weyl Hamiltonian   on the surface $x\times y$.
 For  a crystal-face which is cylindrical,  any point on the cylinder  makes an angle of   $\phi$  with  the $z$ axis.

\noindent 
We consider a situation where the surface perpendicular to  axis $z$  (the direction of the mass gap is   given by  $H_{\bot}$) is is a surface of a cylinder. Any point on the surface can be viewed as rotated  by an angle $\phi$ (the  new ax    $z'$ makes an angle   $\phi$  with the original axis $z$ for which the mass gap has been introduced ) The axes $z'$  becomes the radial direction on a cylinder.  
 The  transformed  Hamiltonian $ H^{'}_{\|}$  is expressed in terms of the covariant derivative  $\nabla_{a}$. $ H^{'}_{\bot}$  and  $H^{'}_{\|}$  are given by  :
\noindent
$H^{'}_{\bot}=\tau_{3}[-m_{0}(\frac{r}{\cos(\phi)})-m_{2} (\frac{r}{\cos(\phi)}) \cos^2(\phi)\partial^2_{r} ]+[i\tau_{2}\cos(\phi)-i\tau_{1}\sigma_{x } \sin(\phi)\partial_{r}]$
\noindent
$H^{'}_{\|}=(-i)[\tau_{1}\sigma_{x}\cos(\phi)(-\nabla_{2})+\tau_{1}\sigma_{y} (\nabla_{1})]\equiv i\tau_{1}[\sigma_{x}\frac{\cos(\phi)}{r}\partial_{\phi}+ \sigma_{y}\partial_{x}+\frac{\sigma_{z}}{2r}]$

\noindent
 The Hamiltonian $ H^{'}_{\bot}$  has  zero mode solutions  $ \hat{\Omega}(\vec {r})$, 
$ H^{'}_{\bot}\hat{\Omega}(\vec{r})=0$   where      $\hat{\Omega}(\vec{r})=\eta_{\tau,\sigma}\Omega(\vec{r})$  is given as a product of  a scalar function  $\Omega(\vec{r})$   and  the spinor $\eta_{\tau,\sigma} $ . The  scalar function is localized at $r=L$  and vanishes for  $ r  \rightarrow\infty$. Using the eigenvalues of $\sigma^{x}$ and $ \tau_{1}$   we find:

\noindent
$\eta_{\tau=1,\sigma; cos(\phi)>0}=\frac{1}{\sqrt{2}}[1,e^{i \sigma\phi}]\otimes\frac{1}{\sqrt{2}}[1,\sigma]$; 
$ \eta_{\tau=-1,\sigma; cos(\phi)<0}=\frac{1}{\sqrt{2}}[1, -e^{i \sigma\phi}]\otimes\frac{1}{\sqrt{2}}[1,\sigma]$ ;  $\sigma=\pm 1$

\noindent
  Using the  eigenvectors      $|1>=\frac{1}{\sqrt{2}}[1,\sigma=1] $  and   $|-1>=\frac{1}{\sqrt{2}}[1,\sigma=-1]$  we introduce the rotated   Pauli matrices   $S_{3}= \frac{1}{2}[|1\rangle \langle 1|- |-1\rangle\langle-1|]$,  $S_{1}=\frac{1}{2}[ |1 \rangle\langle-1|+  |-1\rangle\langle+1|]$  and      $S_{2}= \frac{1}{2}[-i|1\rangle<-1|+i  |-1\rangle\langle+1|]$ for the orbital part 
\noindent Using the eigenstates $\eta_{\tau=1,\sigma; cos(\phi)>0}$ we compute   the projections   $<\tau_{1}\sigma_{x}>|_{\tau=1}=S_{3}\cos(\phi)$ ;$<\tau_{1}\sigma_{y}>|_{\tau=1}=(-S_{2}\cos(\phi)+S_{1}\sin(\phi))$ 
 and find the Hamiltonian for each point $\phi$ :  
\noindent
$ H^{\cos(\phi)>0}_{\|}=(-i)\Big[-\frac{cos^2(\phi)}{r}S_{3}\partial_{\phi}+ \cos(\phi)
(S_{2}-S_{1}\sin(\phi))\partial_{x}-\frac{1}{r}( \cos(\phi)S_{1}+S_{2}\sin(\phi))]
=-H^{\cos(\phi)<0}_{\|}$

\noindent
 We observed that the rotation of the crystal-face for a cylinder is different from the surface state of  a cylinder. In particular we mention the   change sign Hamiltonian    for a  large angle, $\cos(\phi)<0$.  We will compute the eigenvectors for eq.$(12)$ and  determine  the  surface properties  such as  spin texture,  and surface currents  which  affect  the  Photoemmision,  the photoconductivity and the  optical conductivity. 

\vspace{0.2 in}

\noindent
\textbf{Conclusions}
\noindent
To conclude the method of curved spaces in momentum space has been introduced. The method has been used to derive topological invariants from response theory.
The theory has been applied to topological insulators topological  superconductors and persistent currents. We have also  studied the effect of the curved surfaces on the photoemission spectrum.

\end{document}